# Low Reflectance All-Glass Metasurface Lenses Based on Laser Self-generated Nanoparticles

*Jae Hyuck Yoo, Nathan J. Ray, Mike A. Johnson, Hoang T. Nguyen*
*and Eyal Feigenbaum\**

Lawrence Livermore National Laboratory, Livermore, California 94550, USA

E-mail: feigenbaum1@llnl.gov

**Abstract:** Optical metasurfaces, comprised of subwavelength nanostructures, hold a great promise to high-power laser optics but also a limited pertinence due to their currently limited aperture size, throughput and durability. Here, an alternative approach is presented, reliant on laser-controlled self-organizing mask formation followed by ion etching which results in an all-fused-silica-glass metasurface. Two 1 mm diameter optical elements (an axicon lens and a shadower) are fabricated and their optical performance is validated at 532 nm wavelength with an extremely low broadband reflection (<0.15%) - a result of the unique metasurface elements shape. The self-organizing working principle enables producing large amounts of nano-elements at-once, thus a path for aperture scaleup. It also enables generation of sub-100 nm nanoelements, thus a path to short wavelengths operation. Two key advancements towards viability are presented: a laser scan with in-situ transmission feedback enables patterning the etching mask to a prescribed nanoparticle distribution, and a crafted beyond-mask-erosion-point etching of the mask enables increasing the metasurface phase difference to at least $\pi$, while keeping extremely low reflection across it. This paves a path to high-power lasers optics, requiring large aperture, high throughput and laser light durability.

## 1. Introduction

High power and energy laser (HEPL) systems have been rapidly growing and paving revolutionary technology paths, wherein metasurface (MS) technology could become a key enabler of further potential growth. Some potential implications of this field include compact particle and x-ray radiation sources for medical and civil engineering, fundamental high energy science, and directed energy for communications and defense [1-5]. One application that recently captured the global community's attention and imagination is laser-based thermo-nuclear fusion based on inertial confinement fusion (ICF) with the potential to unlock fusion energy as a reliable, sustainable, and safe solution for clean energy sources. The recent repeated successful achievements of fusion at the National Ignition Facility (NIF) have established the scientific feasibility of the laser-driven ICF approach and further spiked the interest in this technology [6-8]. MS could advance HEPL system in terms of size-weight-and-power (SWaP), critical to their operation. Large apertures meant to spread the power for damage reduction drive bulky optics (e.g., NIF wedge focus lens has aperture of about 40 by 40 cm and regions that are more than 4 cm thick)[9]. A critical observed damage mechanism is filamentary damage in the bulk which depends on intensity and length path in the glass[10, 11]. By reducing the thickness of the optics using MS, the allowed power before damage could be increased substantially, since this damage formation intensity onset is inverse proportionally dependent on the optics thickness. MS also offer benefits of reduced size and wight, as well as potentially increased throughput resulting from enhanced flexibility in wavefront tailoring.

However, the optics on HEPL systems are required to be resilient to laser-induced damage in addition to large optical apertures (tens of centimeters). For that reason, fused silica glass is a material of choice for NIF optics [9]. Since adding other materials and material interfaces could decrease the damage threshold, the approach we pursue is of all-fused silica glass MS. All-glass MS have been demonstrated with a variety of fabrication methods and geometries of their nanoelements, typically, each meta-element written individually by deep subwavelength lithography (e.g., electron beam) [12, 13]. A common challenge with such approaches is the limitations on the possible aperture size of the optics due to the demanding requirements on the allowed tolerances for the meta-elements' geometry. Recently, a fully CMOS compatible processes and materials using deep ultraviolet projection lithography (DUVPL), has been used to demonstrate 10 cm all-glass meta-lens, with satisfactory performance to image celestial bodies [14]. This impressive demonstration paves a viable path to large aperture imaging systems, and perhaps high-power laser systems. However, for HPEL optics requirements some challenges and limitations lie ahead, which could be perhaps better addressed by the alternative method presented here.

We have previously presented the basic process for the alternative approach based on laser-self-generated nanoparticles mask (LSG-NPM)[15, 16]. This 4-step process, as illustrated in Fig. 1a, consists of depositing of an ultra-thin metal layer on the fused silica glass, applying heat with laser to dewet the thin film and generate self-organized formation of metal nanoparticles, and use thereafter the nanoparticle as directional etching mask to imprint it into the fused silica glass (typically, reactive ion etching (RIE)), where the sacrificial metal mask is finally washed away. The locally induced phase delay (PD) by the metasurface on the incoming wave depends on the depth of the MS layer and the effective index of that layer. In this MS type, the effective index is a result of Brugman's mixing and could be loosely described in each depth layer using the refractive index which is the volumetric averaging of the glass features and the air vacancies indices[17]. Therefore, one of the determining factors of the PD is the mask fill-factor, FF, (i.e. masking factor, mean area covered by the mask nanoparticles), and we have previously shown that this FF is controlled by the local temperature induced by the laser irradiation[15]. Since the local laser irradiation on an area controls the mask nanoparticle FF and therefore the local index within that area, this method avoids the individual writing of meta-elements – relying on self-generation process and presenting an inherently size-scalable approach. However, in the previous demonstration [15], only rudimentary shapes have been presented, leaving some critical advancements to be made in order to demonstrate a viable approach for optical elements – as will be presented here.

The CMOS compatible projection lithography (i.e., DUVPL) approach presents advantages and disadvantages compared to the LSG-NPM approach when considering them for the applications of HPEL optics. The CMOS compatibility of DUVPL is based on a proven production process and thus presents an advantage in terms of producibility. It is also based on the shape sensitive MS-elements that enable more control of optical properties. The demonstration in [14] puts it further ahead in terms of maturity level. However, several HPEL-related limitations still remain, that could perhaps be circumvented by the self-generation-based approach (i.e., LSG-NPM): (1) Deep ultraviolet (DUV) machines have a limited stepper size of about 2-3 cm size, requiring reliance on stitching fields, leaving gaps in the MS. These gaps may lead to reduced transmission or to downstream modulations that could result in damage to other optics in the system. Moreover, the available DUV machines limit the near-term attainable aperture size. (2) Substantial transmission variations across the MS due to linkage between the local PD (i.e., effective index) and the local transmission of the elements constituting MS. (3) Unlike the case of classic optics where the reflection is commonly mitigated by applying an antireflective (AR) coating, applying such coating atop MS is challenging. (4) Mechanical stability considerations for HPEL optics could require thicker elements (e.g., when serving as a vacuum barrier), which may be incompatible with reliance on CMOS equipment to reach the needed aperture-size. (5) The DUV resolution may pose a limit on creating metasurface optics (metaoptics) for shorter wavelengths (e.g., NIF final optics operate at the 351nm wavelength). While using e-beam lithography instead of DUV should enable sufficient resolution for ultraviolet metasurface operation, it would present substantial challenges in terms of aperture scalability. LSG-NPM has been demonstrated in the past to address the ability for scaling down the nano-features mean period to few tens of nanometers by reducing the metal film thickness [18]. Furthermore, laser-induced damage robustness in fused silica glass is achieved in the NIF by applying a specialized etch based process to remove the surface Bielby layer [9, 19, 20], which LSG-NPM was previously demonstrated to be compatible with [21]. As will be demonstrated here, in the LSG-NPM case, the meta-elements are designed to taper away from the substrate such that the reflection becomes negligible, addressing the challenge of applying an AR layer atop the MS.

In this work we are advancing some substantial knowledge gaps of the LSG-NPM approach. Past works shown that rudimentary shapes with relatively uniform effective index within them could be generated[15]. Furthermore, very fine control of the meta-features height-profile shape (e.g., sidewall sloping, height, spacing, etc.) with this method have been demonstrated[16, 22], as well as its laser induced robustness[21, 22]. However, for a viable metaoptics approach, a predetermined and smoothly crafted PD shapes capability should be demonstrated. Due to the previously observed process nonlinearities[23], especially when multiple overlapping exposures raster scans are involved, it was unclear if and how the FF of a mask could be tailored to a prescribed shape. Furthermore, the mask erosion while etching[16, 17] leads to a limitation on the achievable MS features depth and therefore the achievable PD. As will be described here, the simplified concept of operating without exceeding mask erosion is found to limit the PD, and a new method is developed to reach adequate PD for optics implementation. This method is developed based on the study of the PD nonlinear dependence on the etching time and is utilized to demonstrate a similar PD range to the one presented in [14] with the use of DUVPL. We utilize this method to implement two optical elements useful for lasers systems, an axicon focusing lens and a defocusing lens that could be used for casting a shadow downstream, and validate their optical downstream function and their low broadband spectral reflection.

## 2. Generation of nanoparticle masks

Obtaining a nanoparticle mask with a spatial distribution having a FF that tracks a goal prescription requires a method that could overcome the fabrication nonlinearities and nonidealities. The spatially varying laser illumination could be achieved, for example by a raster-scan of a small beam or by shaping a beam that has an aperture larger than the optical element. The latter is faster as no beam scan is required and does not have to account for complex effects of overlapping beam scans. However, its scalability is limited by the power of the illuminating laser and its resolution by the beam shaping element (e.g., number of pixels of the beam shaping element) – and thus we select in this study to use the small beam raster approach. Past studies revealed that the laser beam power could be used as a knob in varying the resulting mask FF[23], while there is only low sensitivity to varying the dwell time, and therefore in this work we vary the laser power to control the resulting FF. When using raster scan, the amount of the overlap between the different beam paths is a critical design parameter: too large an overlap would leave FF modulation (at the reduced exposure valleys between the raster paths), while extensive overlap would increase the scan time and the nonlinear behavior that is previously observed for multiple exposures at a given point. Moreover, since the predominantly determining parameter of the end-result mask FF is the local induced temperature during dewetting, and that temperature depends on the multiple parameters, some are dynamically changing throughout the dewetting process (e.g., the absorption profile changes dynamically through the interaction with the pulse as dewetting occurs and change the morphology [23]) the relation between the end-result FF and the applied illumination is hard to theoretically predict. To overcome this, we have devised an iterative approach that converges to a desired prescribed shape. One challenge to such an approach is obtaining an in-situ diagnostics information: measuring the FF distribution after each iteration using a method that can resolve the nanoparticles, such as scanning electron microscopy (SEM), would render this iterative approach impractical. Fortuitously, as proposed and demonstrated in [15], optical transmission measurement could be used as a proxy for the mask FF when using an illumination wavelength that is not in the spectral resonance of the mask nanoparticles. At such a wavelength, the transmission simply represents the geometrical obscuration of the nanoparticles to the incident light, governed by the mask FF. This enables a relatively simple and quick in-situ scanning of the FF by measuring its transmission proxy and thus enables correcting the laser scanning power pattern in order to converge to the prescribed nanoparticle mask shape.

An iterative converging mask patterning setup based on transmission measurement as in-situ feedback was developed, showing efficient conversion and relatively quick fabrication for a prescribed transmission profile. The principal process is illustrated in Fig. 1b. We implement a radial laser scan with the laser beam power being a function of only the radius, adequate for the azimuthally symmetrical optics implemented in this study (e.g., lens). An initial prescription of power as a function of radius is applied, $P(r)$, illustrated as radially linear with negative slope, in the Fig. 1b illustration. The prescribed radial power exposure results in a nanoparticle mask which then is back illuminated to obtain its transmission. The radial transmission, $T(r)$, is then derived from the measurement as the mean transmission value at a radius. The resulting $T(r)$ is then compared with the prescribed $T(r)$, resulting in the corrected $P(r)$ for the next iteration.

In the mask patterning setup, the sample is loaded on a stage that could traverse between the laser patterning station and the transmission measurement locations. The mask patterning is conducted with a 5W continuous wave (CW) 532 nm wavelength laser. We note that we expect this approach to be applicable when using other wavelengths, since the underlying mechanism is laser-induced temperature[15], and since we observed similar dewetting behavior with 1064 nm CW laser[23]. The beam power is computer-controlled using a rotation of a half-wave plate (HWP) sandwiched between two polarizers, and the beam scan is controlled by two axis galvanometers. The scanning pattern was set to be spiral to avoid radial transitions artifacts. The constant time per revolution was set for ease of automation where the increment of the linear speed with the radius is enabled by the low sensitivity to the dwell time parameter. After the beam is directed by the two galvanometers, it is focused on the sample using a focusing lens (f=100 mm) producing beam spot of 8.5 µm on the sample. The spiral radial pattern was implemented with constant radial speed of 10 µm/s (i.e., under 9 minutes for a 1 cm diameter pattern at that scan speed), rotation speed of 720 degree/s (i.e., 2 revolution per second), and radial scans gap of 5 µm (which was found to result in a smooth transmission profile and unnoticeable transmission modulations). In the future, the process time could be further reduced by maximizing the scan speed for each radii and the beam size. After the mask was patterned the stage moves to the transmission measurement position and using back illumination of 405 nm wavelength LED the transmission measurement is conducted. A camera is then used with objective lens allowing for the patterned mask to be characterized (0.3545 pixel/µm resolution). A more detailed description of this setup is given in the supplementary (in Fig. S.1.).

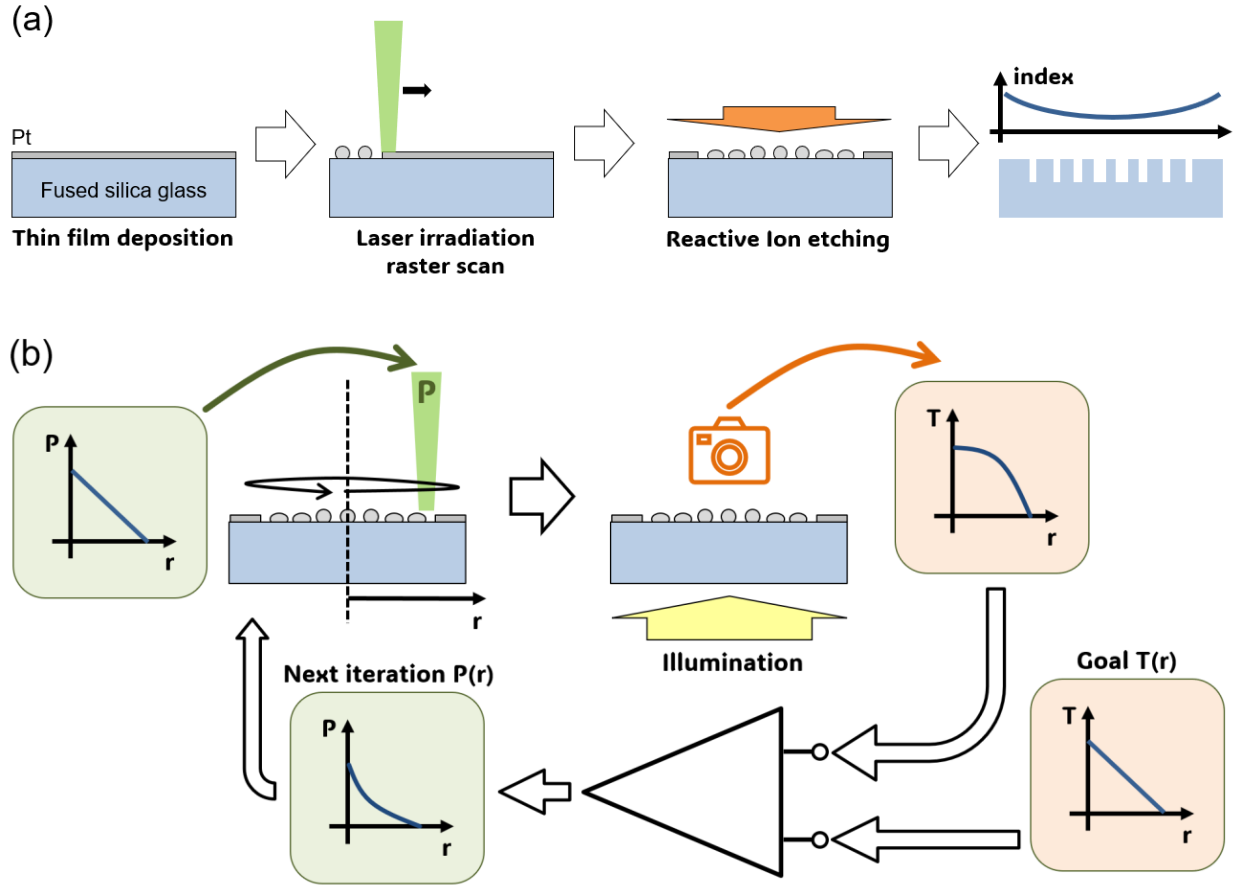

**Fig. 1. Schematic illustration of process:** (a) the laser-self-generated nanoparticles mask (LSG-NPM) approach description to generating MS; (b) The iterative approach for obtaining a prescribed transmission profile of nanoparticles mask with radial laser scan and in-situ measurement of transmission.

This mask patterning process is notionally illustrated in Fig. 2a for a case where the goal is obtaining a linearly growing radial transmission. Since the transmission is known to monotonically follow the applied laser power, we use as an initial radial power guess that grows linearly with radius, to match a linear growth in the prescribed transmission. Since the radial scan starts from the center outwards, it translates to a linearly growing function as a function of scan time (the red curve in Fig. 2a represents the HWP angle controlling the transmitted laser power in arbitrary units). The resulting radial transmission of the sample is growing radially as intended, but in a nonlinear relation (black curve). This allows construction of a look up table (LUT) relating the laser power parameter and the resulting transmission (green curve) – assuming the power applied at each radius is locally resulting in the transmission at that location. For the prescribed transmission function range and shape, we can utilize the inverse LUT to predict the power as function of time to be applied at the next iteration of the process. The new iteration is applied at a new location (not on top of the previously patterned area) – resulting in this hypothetical example in the desired radial transmission. In practice, as we will present hereafter, nonlinearities will sometimes result in some residual imperfections and require additional iterations. Once a temporal power curve has been developed, it could be used in the future as a recipe without the need to repeat this convergent development process. This could present time saving implications for situations involving replications, since once the P(r) curve has been developed it could be applied to manufacture multiple such elements without the need to rederive that formula (e.g., production of multiple lenses of same design).

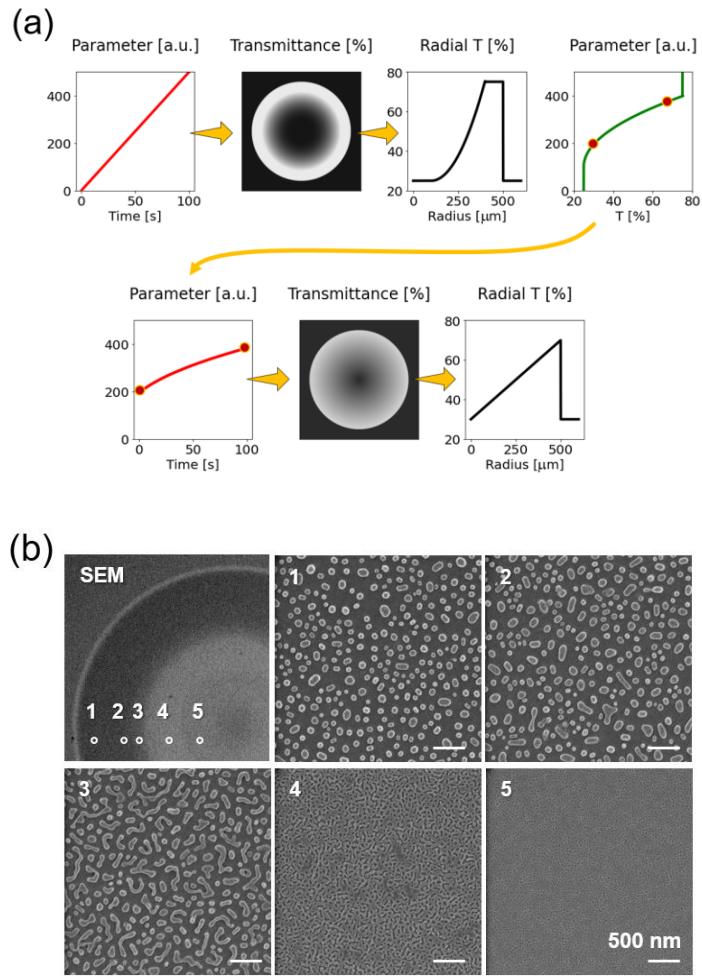

**Fig. 2. Prescribed transmission profile process with radial laser scan:** (a) detailed example of the iterative printing process resulting in a prescribed radial transmission profile; (b) SEM of end-

resulting mask, and enhanced magnification to illustrate the mask distribution radial change (scale bar length – 500 nm).

The end-result patterned mask following the laser-patterning process, as detailed above, is characterized by SEM, as presented in Fig. 2b. As expected, enhancing the power radially results in enhanced transmission due to the reduction in the observed nanoparticles mask FF (e.g., the highest FF is observed in region '5', which was exposed to the weakest irradiation). In this example the mean period obtained is 241 nm (based on spectral analysis of the SEM), which should suffice for the operation wavelength of the metasurface optics in this study, being 532 nm (i.e., at which we will test the optical performance). However, as previously observed [15, 18] the resulting period could be controlled with the process parameters, and could be further reduced, if needed e.g., using a thinner deposited metal layer to match even shorter wavelengths (throughout this work a 7 nm thick layer of Pt film has been used).

We tested the above mask patterning method for two radial linearly varying transmission prescribed profiles: ascending and descending. The choice of linear transmission profile was made to simplify the visual illustration of converging to the prescribed slope. The ascending and descending slopes are chosen as they demonstrate two key optical functions: focusing and defocusing. As will be detailed later, even though the developed process resulted mostly in a linear mask transmission profile, nonlinearities of the etching process resulted in a nonlinear profile of the MS effective index (i.e., PD). Nevertheless, the radially ascending and descending effective index represent useful optical components: a conic shadowing element (also known as shadow cone blocker, SCB[9, 24]) and a focusing element, accordingly. The SCB elements are used in laser systems to suppress further laser induced growth of damage sites downstream of the SCB, by casting a shadow over the damage site. These elements function by deflecting the light incident on the intended aperture outside of the beam path (differently than, for example, the shadow forming downstream from a location of phase singularity created by an orbital angular momentum (OAM) phase-screen [25]). Radial descending linear and parabolic profiles would result in an axicon and a lens, both advantageous for a variety of applications of material processing and imaging. While the axicon forms a long focal range - beneficial for reduced focal distance sensitivity, the standard lens generates tighter axial focusing. Therefore, we chose implementing the radially ascending and descending nanoparticles masks profiles, which have both useful and intuitive functions to better illustrate the approach. In addition, they represent the two basic building blocks to furthermore complex optical elements. For example, a diffractive optical element, being key to reducing the f-number of lenses, would require alternating radial segments with ascending and descending index slopes.

The processes of laser-patterning nanoparticle mask with linear ascending and descending radial transmission are illustrated in Fig. 3. In Fig. 3a, a conversion to a radially ascending radial transmission is shown to be obtained within only two conversion steps (each step represented by a row in this sub-figure), where one step is defined as a one full cycle through the steps represented in Fig. 1b. After laser-patterning, using an initial guess of linear radial increment of the HWP angle as a function of the scan time (similarly to the example in Fig. 2a), a nonlinear behavior of transmission vs radius is observed. In this example, the HWP angle prescribed profile was chosen arbitrary to pertain a linearly radial change between 40% and 55% over 0.5 mm radius. Based on the radial transmission a LUT is used to predict the next HWP profile to be used. Using the suggested HWP angle, the measured radial transmission closely meets the prescribed profile. Even though there are some small noticeable deviations from linear, we decided that such visibly negligible level of deviation is adequate for this evaluation. For the radially descending transmission case that is illustrated in Fig. 3b, 3 steps were required for convergence. We have used in this case as an initial guess a radially descending HWP angle to better match the prescribed profile.

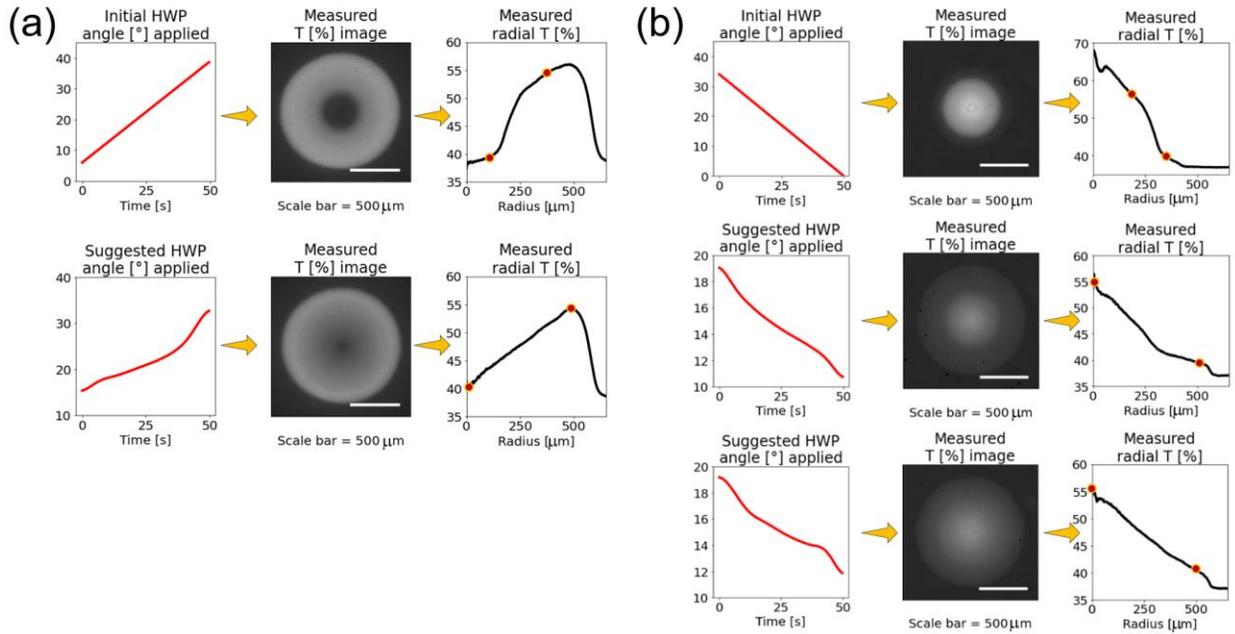

**Fig. 3. The processes of laser-patterning nanoparticle mask with linear radial transmission: (a) ascending and (b) descending:** each conversion step (one whole cycle through steps in Fig. 1b) is illustrated in a separate line, including its used HWP angle vs time used for the laser patterning, the measured transmission image, and the resulting radial transmission.

### 3. Reactive ion etching (RIE) of the NP mask to form a MS

For a patterned nanoparticle mask, the end-result metasurface effective index could be made proportional or inversely proportional to it, based on the RIE process length. Fig. 4a illustrates schematically the difference between the two cases of short and long RIE time. In the short etching case, some portion of the mask remains after the RIE over the entire patterned area. Therefore, the formed metasurface layer depths are equal across the structure, and the columnar glass area follows the mask FF. In this case, regions of small FF (and transmission, where higher laser intensity was used for dewetting) will result in a smaller PD of the end-result metasurface. In the long etch case, at some of the segments the nanoparticles mask is completely eroded and disappear during the RIE process. Even though the mask material has substantially better erosion resistance than the substrate, it is still finite (the Pt mask used here erodes about 30x slower than the glass substrate, depending on the RIE recipe and nanoparticle mask parameters). Observations show that for a given initial film thickness, the dewetting-resultant NPs in lower FF segments are also taller and will remain longer through RIE process, which is consistent with first principals of mask material volume conservation. Once the mask disappears from a region, the structure will be etched conformally, for idealized parallel ion trajectory milling. In practice, angular RIE rays' distribution, nonuniformities and nonidealities will devastate the structure as the RIE process becomes exceedingly longer. However, in an idealized manner, as illustrated in Fig 4b, for a long RIE process segments with small FF will result in deeper metasurface pillars, whereas larger FF segments will result in shorter ones. Both the increment in the metasurface FF and its depth contribute to the increase in its PD, and since, relative to fused silica, the Pt has such a large erosion durability when the RIE becomes long enough the local MS depth dominates the resulting PD over the FF. Therefore, the level of control of the MS PD is so substantial at the RIE step that it could potentially alter the nature of the resulting device from being a focusing to becoming a defocusing MS optics. Even though the short etch process simply transfers the patterned mask to the substrate it has few key disadvantages. First, it results in a smaller PD across the structure, since the longest possible etch time is dictated by the depth of the as-deposited layer (being the shallowest region of the mask). Another limitation is the varying reflection across the sample. For the short etch process, the MS pillars typically have close-to-vertical sidewalls and therefore a uniform effective index into the layer. Since the reflection of a surface layer is sensitive to its refractive index (and could vary harmonically between perfect impedance matching conditions to the bare surface reflection [26]), for such a MS with a uniform thickness and varying effective index across the layer finding reduced reflection parameters is extremely challenging. Alternatively, in the long etching case, the end resulting MS features will be elongated features tapering from a broader base near the substrate to a narrow tip at the air interface. This result in a tapered effective index into the MS depth and thus a reduced reflection insensitive to the index profile and depth. Therefore, in the long etch case we could achieve a larger PD and with lower reflection across the metasurface, making it preferable.

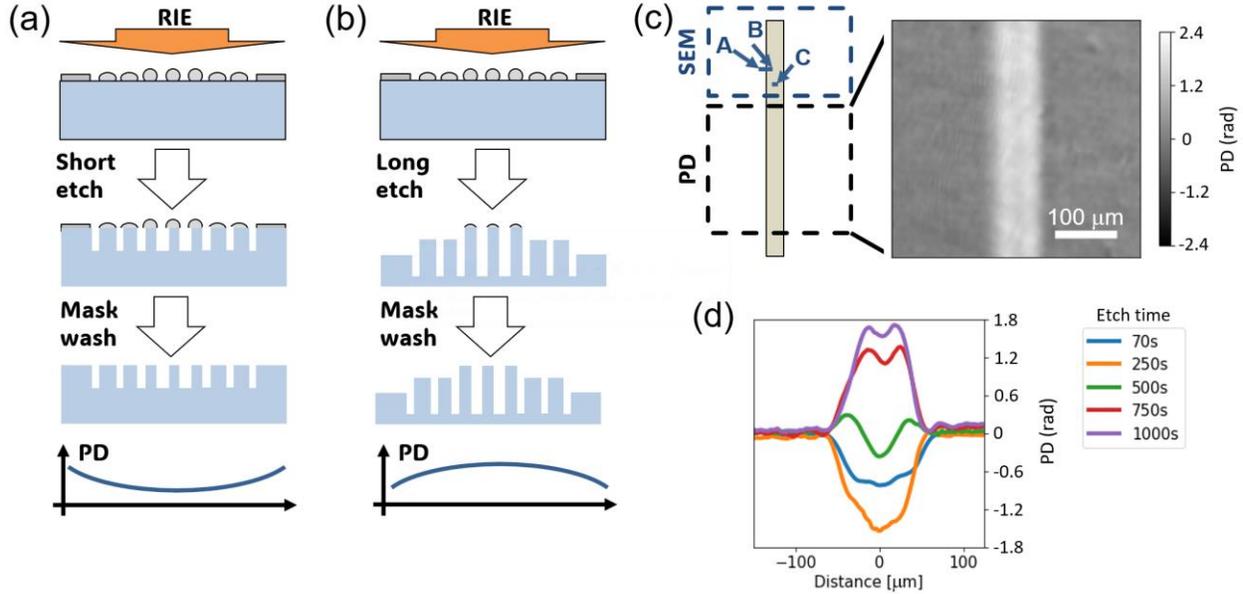

**Fig. 4. Etching time effect on the end-result metasurface:** a schematic illustration of the (a) short vs (b) long RIE etch time cases and the resulting phase delay (PD) profile; Laser scan lines study with varying etch times: (c) geometry schematics and PD measured using PSDI for 1000 s etch time, and (d) PD measurement cross-section for different etch times.

To test the etching effect on the resulting PD we have tested varying RIE time and examined the SEM and PD of the resulting structure. To simplify the characterization, we have generated a laser line scan made over a masked fused silica substrate (also 7 nm Pt layer) at 1 mm/s scan speed. Followingly, we cleaved the sample to cut the line scan into 5 separate sample pieces to be RIE processed with a different etch time: 70 s, 250 s, 500 s, 750 s and 1000 s, for each of the samples (where $C_4F_8$ supplied the reactive ions), and followingly the remaining mask was chemically removed. The characterization of the samples is illustrated in Fig. 4c. The linear scan is represented by the tan elongated rectangle, where one section was utilized for SEM analysis, and another for PD measurement – as denoted in the figure. The optical wave-front phase difference locally induced by the MS modified areas is measured using a phase shifting diffraction interferometer (PSDI) technique, at $\lambda = 532.2$ nm free-space wavelength [27]. The PD is then averaged along the scan direction, for noise reduction, and the end-result for the different samples is presented in Fig. 4c. For short etch conditions, observed for the 70 s and 250 s etching time cases, an increase in the laser irradiation (higher towards the scan center) resulted in a lower mask FF and therefore more negative PD when compared to the non-processed region. The PD changes monotonically from the scan center – following the laser irradiation. The PD range achieved with the 250 s is greater than that with the 70 s one, as expected, getting to almost ¼ a wave (consistent with the values obtained by dividing the film thickness by the FF values range and multiplying the resulting particle height range by the mask-substrate etch ratio). As the etching time increases to 500 s, the transition to the long etching regime is observed, where the two competing contributors discussed earlier (FF and MS depth) factor-out each other and reducing the mean PD across the structure. For longer etches, 750 s and 1000 s, the long etching conditions are observed. The PD across the structure increases for longer etching time, exceeding ¼ wave for the 1000 s conditions (i.e., PD > $\pi/2$). We note that the competition of the contributing conditions results in a non-monotonic behavior, and the formation of a notch at the center of the scan line. Therefore, for the patterning components, we will be using the mask FF range that for these etching conditions result in the monotonic change range observed in between about 25 µm and 50 µm distance from scan center. Choosing of a section of the PD vs mask FF with the largest monotonic PD transition, tailored for the chosen etch depth, ensures a monotonic and thus more controllable process of the overall laser patterning of the metasurface PD.

The SEM cross sections of the different etch time conditions of the scan lines are depicted in Fig. 5. As illustrated in Fig. 4c, an SEM section at the center of the scan line is denoted by "C", whereas an SEM section at the edge of it is denoted by "A-B", where B is closer to the scan line center. For the very short etch conditions of 70 s RIE time, the expected vertical side walls of the MS nanofeatures are observed, with reduced FF towards the scan line center, and uniform MS depth across the structure (black dashed line are used to guide the eye). The high FF at "A" outside the beam track (i.e., seeing the beam pedestals) follows the mask structure presented in Fig. 2 at very low irradiation (region "5"). For the 250 etching conditions, we observe at the center cylinder-type shapes expected for short etching conditions, where the mask nanoparticles are the tallest, and last longer under etching. Yet, at the edge "A-B" section, we could detect a small depth change associated with approaching the longer etch conditions (yellow and white dashed lines are added to guide the eye to the MS top surface in "A" and "B", correspondingly). Therefore, we can interpret it as the mask at point "A" has been fully eroded after 250 s. The SEM of the etching conditions at longer etch times present the much taller and conic-like MS nanofeatures anticipated, where the MS depth varies significantly across the scan line. While the mean periods observed in "C" region for the 500 s and 750 s case are similar, the depth is clearly larger in the latter, as expected.

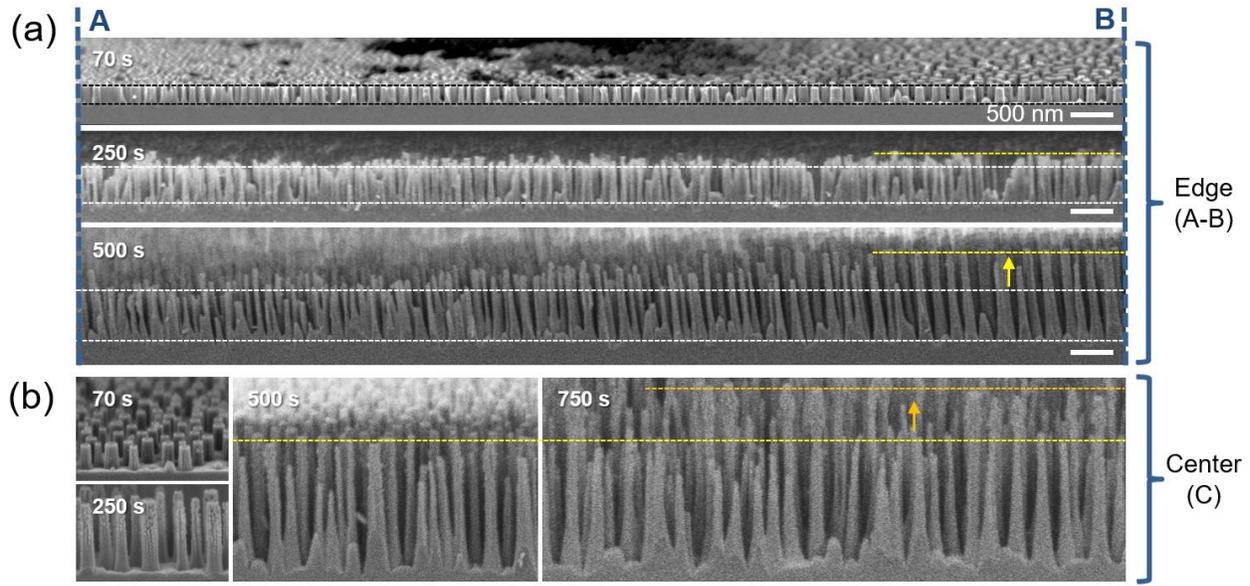

**Fig. 5. Etching time effect on the end-result MS geometry:** Cross-sectional SEM for different etch times (detailed in s on each of the SEM images) at scan line presented in Fig. 4c: (a) edge of scan line ("A-B"), and (b) at its center ("C"). dashed lines are given to serve as a reference line and guide the eye to the differences in height across the sample.

## 4. Fabrication and characterization of optical elements

The laser-patterned masks with the linear-radial descending and ascending transmission profiles were etched at the optimized RIE conditions to produce an axicon lens and a SCB components, accordingly. The mask laser-patterning procedure described in Fig. 3 was applied on a sample (7 nm of Pt coating on fused silica, as before) aiming at a linear radial prescribed transmission profile of the NP mask varying between about 40% to 60% (to avoid confusion, these transmission values are of the metal NP mask and are part of characterizing its FF pattern. Differently, the end-result metasurface is obtained after etching and mask-removal, leading to much higher transmission values and its characterization is discussed later in this section). For the ascending profile (SCB mask) we have extended the value at the outmost radius (i.e. at 0.5 mm) to a radial range outside it (i.e., constant mask FF between 0.5 mm and 1.5 mm) to form some constant surrounding region at the maximal transmission value obtained at element maximal radius. The resulting profiles (shown in the supplementary in Fig. S.2) follow the linear radial slopes, with more noticeable deviation at the central 100 μm section. The axicon mask resulted in a larger transmission range than the one of the SCB (about 35%-65% vs 45%-60%), as a result of the central region imperfections and of the axicon mask benefitting from additional continuation drop to the laser-untreated surrounding area. As will be shown later, this will result in a larger PD for the axicon than the SCB, and since it is less critical for the function of the SCB element in the context of this work we have not further optimized the SCB mask. The 1 mm diameter axicon mask and the 3 mm SCB mask laser-manufacturing times were only about 3 and 1 minutes, accordingly, and contained an estimated number of nano-features of about 120 millions and 13 millions, accordingly. This is smaller than the 18.7 billion nano-features made on the 10 cm record large metasurface in [14] – still, size scalability growth potential of this method is promising and topic for future exploration. Furthermore, the linear dependence of the processing time on the radius was made arbitrary for convenience and could be further optimized in the future to reduce the mask manufacturing time.

RIE optimization studies of the two elements' masks were conducted, with maintaining the same etching conditions for both elements, resulting in the two metasurfaces optical elements (near the 1000 s conditions described earlier in Section 3). The SEM images of the axicon and the SCB are shown in Fig. 6a and 6b accordingly. The diameter difference between the two metaoptics (about 1 mm vs 3 mm) is the result of the additional segment of the SCB outside the varying index segment (as discussed earlier). The noticeable dark spot at the center of the axicon (Fig. 6a) could be traced back to an artifact in the mask transmission (see Fig. S.2). This ~20 μm radius center deviation is a mask-patterning artifact due to laser scan sequence and could be addressed in the future by adjusting the dwell time of the laser at the center. Two zoomed-in SEM on two sections of the SCB are given to show the change in the metasurface nano-elements, where the main change observed is in their length, consistent with the long-etch design (as described in the previous section). More zoomed-in SEM Sections are brought in the supplementary in Fig. S.3.

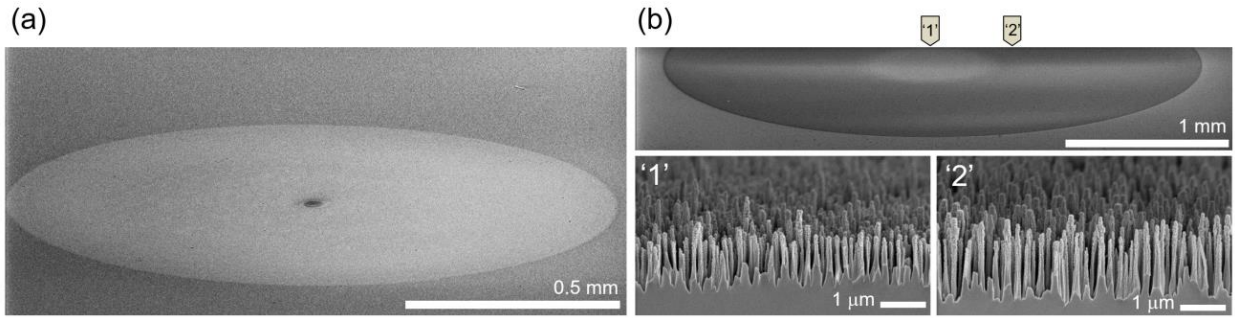

**Fig. 6. SEM of the two metaoptics (MS optics): (a) axicon lens, and (b) shadow cone blocker (SCB)**, with two enhanced magnification cross-sectional SEM of sections indicated by '1' and '2'.

The PD of the two optical elements is measured using PSDI at 532 nm wavelength and is showing the anticipated profile of a focusing and defocusing lenses, where the PD for the axicon presents about π range. The PSDI measurement is used to directly characterize the phase screen applied by the metasurface to the incident beam. The Axicon phase screen resulting from the descending mask transmission profile is shown in Fig. 7a and presents a π rad phase range (similar to the range demonstrated using DUVPL for a lens in [14]). The resulting phase screen is azimuthally symmetric, smooth, and radially monotonic - except for the noticeable point at the center (traced back to the mask transmission artifact, as discussed earlier in relation to its SEM image). Fig. 7b presents the PD cross-section, presenting a close resemblance to a linearly radial profile of an axicon, except for the anticipated deviation at the central section that could be traced back to the mask imperfections (~100 μm radial), and perhaps also to etching nonlinearities – a potential topic for future studies. The phase screen and cross-section for the radially ascending mask patterning of the SCB are presented in Fig. 7c and 7d, accordingly. Similarly to the axicon case, a close to linear radial dependence is observed outside the central ~100 μm region, that is noticeable and likely resulting also from the mask transmission. The π/2 rad PD range, being smaller than the axicon case, could be mainly attributed to the smaller transmission range of the mask.

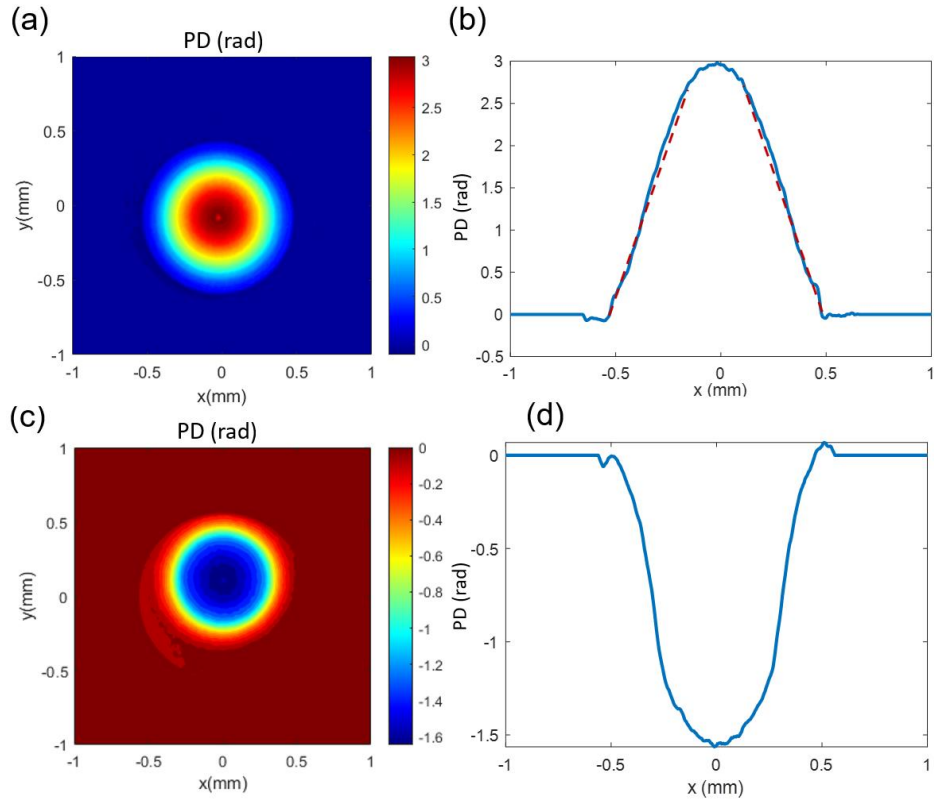

**Fig. 7. Metaoptics PD distributions:** for Axicon MS – (a) PD distribution and (b) cross-section of PD, and for SCB MS – (c) PD distribution and (d) cross-section of PD. The linear dashed lines on (b) have same slope angle and are to guide the eye.

The MS axicon optical function as a lens was demonstrated at 532 nm wavelength using both beam propagation test and simulation. The MS axicon was centered in front of a continuous wave near-Gaussian laser beam, with 0.55 mm waist (slightly larger than the MS) and the intensity profile was measured in planes along the beam propagation (in 10 cm steps, up to a meter distance, where the native beam divergence in this range was verified to be negligible). The intensity distributions in some representative planes are given in Fig. 8a, with the propagation distance from the MS given at the header, and where the excitation distribution is at

z = 0 m (the noticeable periodic linear artifacts are from the camera readout). The tight focusing of the beam after 20 cm, followed by slow divergence, prototypical to lensing of a Gaussian beam, is evident – validating the demonstration of a metalens element function. Beam propagation simulation using the measured PD of the MS (shown in Fig. 7a) was conducted and compared to the beam propagation measurements, to further validate the PSDI-measured PD of the MS and the lensing function of the element. The simulation results given at the same planes as of the measurements are presented in Fig. 8b, showing very good agreement with the measurements, in predicting the beam size evolution and even in capturing some beam shape characteristic (e.g., the secondary ring evolving at longer propagation distance). We note that the laser beam excitation used for the simulation is the measured one. The modeled intensity distribution along the propagation direction is presented in Fig. 8c (cross-section along one of the transversal axes) to illustrate the complete beam evolution for this axicon metalens. The agreement between the measurements and the modeling of the full-width half-maximum (FWHM) of the beam along propagation is shown in the supplementary Fig. S.4 (in all measured planes). Further comparison of the beam propagation simulations and post-analysis between the idealized and measured axicons shows a similar optical function, both showing about 79.5% of the power at focus in the first lobe and a similar focal distance of about 6 cm (see supplementary Fig. S.5).

The defocusing MS element, acting as an SCB and casting a shadow downstream, was demonstrated with beam propagation through the optics. The same measurement described above for the axicon was used to characterize the MS SCB beam propagation. The measurements are depicted in Fig. 8c (the excitation is the same as in Fig. 8a) showing that the MS is casting shadow on the beam footprint. The shadow patch is optimal at about 1 m to the 1.5 m region of downstream propagation, and relatively clear of diffraction spots artifacts (e.g., Poisson spots). The simulations brought in Fig. 8d are in good agreement to the measurements in Fig. 8e, and the predicted intensity distribution along the propagation is given in Fig. 8f. The shadow diameter casted downstream of this MS SCB is about 1 mm, which is comparable to the diameters of the SCB in use at the NIF (extending from 0.36 mm to 0.9 mm of diameter)[9]. However, the typical NIF SCB are designed to cast shadow after about a cm downstream, thus for the same operation much larger PD is needed (and likely requiring a diffraction optics profile).

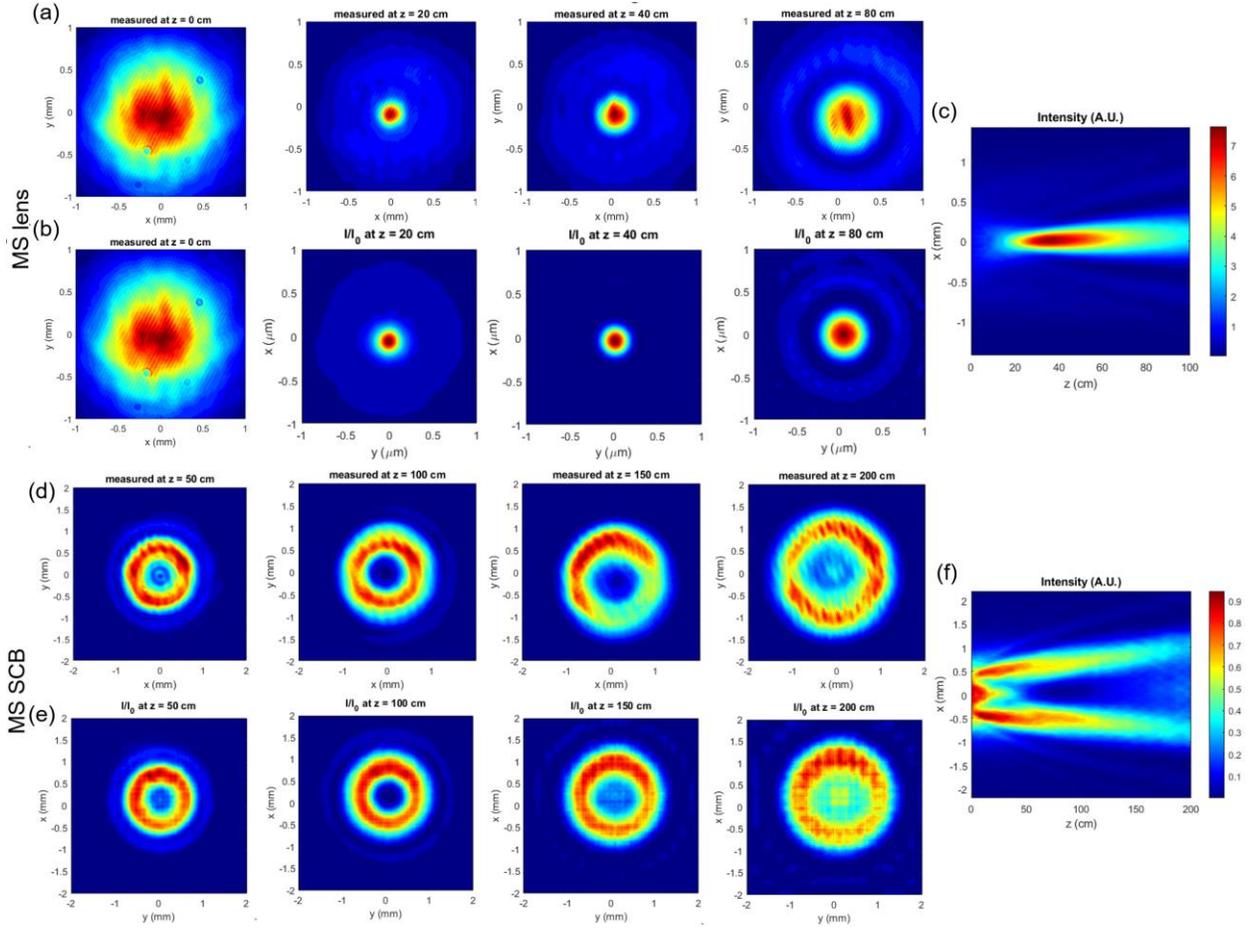

**Fig. 8. Metaoptics beam propogation for axicon and SCB:** transversal intensity distribution at given distances of propagation (given in the title): (a/d) measurements panel (Axicon/SCB MS cases) and (b/e) simulations panel (Axicon/SCB MS cases); (c/f) Intensity distribution cross-section in x (transversal axis) as a function of propagation axis, z, (Axicon/SCB MS cases).

The reflectance of the two MS elements has been measured – showing extremely low and spectrally broadband reflectance. The spectral reflection of the MS was measured at three locations: the two MS elements and a reference area free of optical elements (using Cary 7000 Universal Spectrophotometer), as presented in Fig. 9a. The measurement was done with a square beam spot of about 2.5 mm edge length, slightly larger than the MS elements (i.e., the casting shadow central region for the SCB element case). The reflectance presented is of the surface with the MS (by offsetting the measurement of reference fused silica surface spectral reflection). The measurements show that the reflectance in all three measured locations is substantially lower compared to native fused silica surface. This is especially noticeable for the visible spectrum (relevant to the intended 532 nm operation wavelength of the elements), where the measured sub 0.2% is much smaller than the one for the substrate surface (about 4%). The low reflectance on the reference segment is a result of the MS forming based on the as-deposited mask nonuniformities under the long etching conditions, as observed in Fig. 5 (in segments 'A').

Nevertheless, the infrared spectral reflectance is much improved in regions where the mask has been dewetted by the laser before etching, as the optimal MS depth for AR function increases with wavelength. The reflectance spatial distributions of the MS axicon and MS SCB at 532 nm wavelength are presented in Fig. 9b and 9c, accordingly. A reflectance image calibrated to the spectral measurement at 532 nm wavelength, shows that these elements have an extremely low reflection of less than 0.15%. The two metasurface elements have 95% transmission, with a couple of percent transmission variations around this mean value over the PD range (v. 65% to 95% transmission values range of the metasurface elements used in [14] to cover a π PD range).

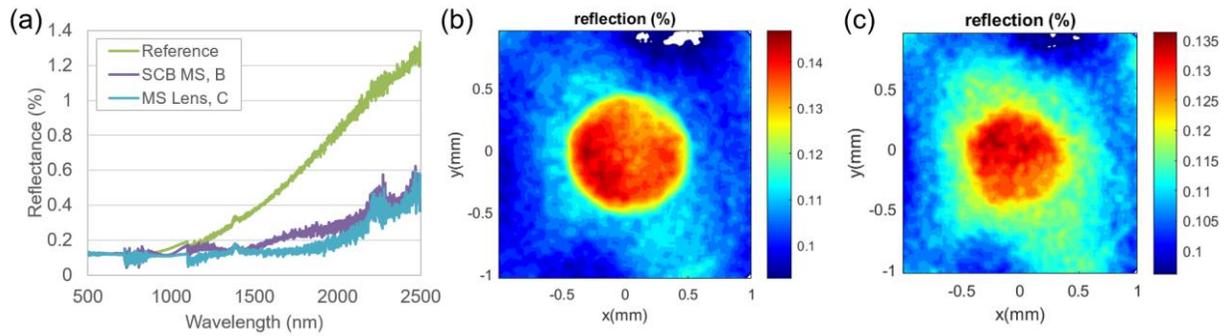

**Fig. 9. Metaoptics reflection:** (a) spectral at the two metaoptics location and location outside the metaoptics area on the sample (averaged on a 2.5 mm x 2.5 mm spot size); reflection distribution of the (b) axicon and the (c) SCB.

## 5. Conclusion

In this work we address two major knowledge gaps that have limited the laser-self-generated nanoparticles mask (LSG-NPM) approach for generating all-glass metasurfaces (MS) from becoming viable. We present here an approach for (1) producing nanoparticle etching mask with a prescribed fill-factor (FF) profile using laser raster scan; and for (2) enabling large enough phase delay (PD) range by optimizing the etching process and the patterning-laser power exposure range. Furthermore, we experimentally demonstrate here for the first time using this approach to manufacture two optical elements having key functions in high power lasers optics with 1 mm diameter: an axicon lens and a shadow cone blocker (SCB).

The LSG-NPM used here is a relatively simple 4-step process based on a widely available and size-scalable equipment: (1) depositing of an ultra-thin metal layer on the fused silica glass, (2) applying heat with a raster scanned CW laser to dewet the thin film and initiate self-organized formation of metal nanoparticles, (3) use thereafter of the nanoparticle ensemble as a mask in a reactive ion etching (RIE) process to imprint it into the fused silica glass, and finally (4) wash away of the sacrificial metal mask. This approach has several potential advantages for the applications of high energy and power lasers (HEPL), in terms of aperture scalability, thick substrates compatibility, meta-elements small enough to fit even short wavelengths in the ultraviolet, low reflectance, and high laser damage and mechanical durability.

One of the obstacles for enabling a viable metaoptics using this approach was obtaining a prescribed and smoothly crafted PD shapes. The challenge results from the complexity and nonlinearities in the mask laser-patterning process. An iterative converging mask patterning setup based on transmission measurement as in-situ feedback was developed, showing efficient conversion and relatively quick fabrication for a prescribed transmission profile. We demonstrated this mask patterning method for two radial linearly varying transmission prescribed profiles, ascending and descending, and showed that it converges to the prescribed profile within a couple of iterations. Some further improvement to this method is likely needed, as some imperfections have been observed and reported (especially at the central section of the mask). The resolution of the mask FF patterning could be refined (and even altered while the process) by changing the beam size. In earlier works the mask fill-factor was shown to closely track the beam shape [15, 23]. Significant variations on a few hundred nanometers length scale when scanning with about 10 micron wide Gaussian beams were observed, suggesting that even more rapid transitions are obtainable with narrower beams. This presents refined patterning resolution that perhaps could be further enhanced with tighter scanning beams, a potential matter for future exploration. To help visualize the application of this mask patterning approach towards diffractive lens design, particularly the necessary sharp transitions between Fresnel zones, we bring in Fig. S.6 in the supplementary an example of such a prototypical manufactured mask. This yet-to-be-optimized mask illustrates the obtainable sharp transitions between the Fresnel zones, and the potential utility for other optical elements, e.g. printing optical grating with prescribed profile and low reflection.

The other major obstacle was that the mask erosion while etching leads to limitations on the achievable MS features depth and therefore the achievable PD. Under the originally proposed paradigm where the etching step simply meant to transfer the mask FF to the substrate, without exceeding mask erosion (i.e., short etch) is found to limit the PD too extensively. By studying the effect of etching length, we found that the PD range for optimized mask transmission range (controlled by varying the processing-laser power) could be substantially increased and still stay monotonic. The end-result MS nano-elements are much longer and tapered shaped, where the length of the features is more dominant in determining the local PD than the FF (opposite to the short etch paradigm). The elongated tapered MS features shape leads to a low broadband reflection. Even though the $\pi$ PD range achieved here is typically considered sufficient for MS optical lens designs, further exploration of additional increase in PD from half to a full wave (i.e., PD of $2\pi$) could enhance the MS optical performance (by e.g., further optimizing the etching selectivity or replenishing the top of the metasurface features with additional mask for further etching, as demonstrated in [17]). Furthermore, it would be beneficial to develop in the future a method for better predicting the connection between the mask transmission and the resulting MS PD due to the etching nonlinearity, to supplement the ability to implement a prescribed transmission etch mask demonstrated here. The converging nature of the process described here is intended to accommodate the nonidealities of the manufacturing process. The mask patterning could be further perfected with more iterations and algorithmic scanning refinements as it has an in-situ feedback, and with incorporating knowledge of the etch transfer function at the intended etch time into the pattern look-up-table (LUT).

We utilized this method to demonstrate two optical elements useful for lasers systems, a focusing lens and a defocusing lens that could be used for casting a shadow downstream and validate their optical downstream function and their low broadband spectral reflection. For this method to be used for large aperture laser systems optics, further scaling up advancements beyond the small metasurface optics shown here are required. While metal deposition and etching machinery with the required scale readily exist, the laser patterning approach presented here is natively scalable by increasing the scanned area radius. While in this study we demonstrated that a sufficient local index form figure is achievable using this method, an optimized laser scan method is likely needed to produce large optics masks with practical manufacturing time (e.g., dynamically varying the beam size and raster speed). Although we have emphasized in this work on designing the optical properties of the metasurface, other properties of the surface could be controlled as well (e.g., hydrophobicity) to enable patterning of both the optical and physical properties of a surface [28].

## Acknowledgements

*Funding.*

Lawrence Livermore National Laboratory (LLNL) Laboratory Directed Research and Development grant (#21-ERD-002).

*Acknowledgements.*

This work was performed under the auspices of the U.S. Department of Energy by Lawrence Livermore National Laboratory under Contract DE-AC52-07NA27344.

19. Suratwala, T., et al., *Chemistry and formation of the Beilby layer during polishing of fused silica glass.* Journal of the American Ceramic Society, 2015. **98**(8): p. 2395-2402.

20. Suratwala, T.I., et al., *HF-based etching processes for improving laser damage resistance of fused silica optical surfaces.* Journal of the American Ceramic Society, 2011. **94**(2): p. 416-428.

21. Ray, N.J., et al., *Enhanced laser-induced damage performance of all-glass metasurfaces for energetic pulsed laser applications.* Applied Optics, 2023. **62**(31): p. 8219-8223.

22. Ray, N.J., et al., *Substrate-engraved antireflective nanostructured surfaces for high-power laser applications.* Optica, 2020. **7**(5): p. 518-526.

23. Yoo, J.-H., et al., *Laser-assisted tailored patterning of Au nanoparticles over an inch-sized area: implications for large aperture meta-optics.* ACS Applied Nano Materials, 2022. **5**(7): p. 10073-10080.

24. Browar, A.E.M., et al., *Laser micro-machining and damage testing of rounded shadow cone blockers on silica glass for arresting laser damage growth by redirection of light.* Optics Express, 2024. **32**(3): p. 4050-4061.

25. Yao, A.M. and M.J. Padgett, *Orbital angular momentum: origins, behavior and applications.* Advances in Optics and Photonics, 2011. **3**(2): p. 161-204.

26. Born, M. and E. Wolf, *Principles of optics: electromagnetic theory of propagation, interference and diffraction of light.* 2013: Elsevier.

27. Guss, G.M., et al., *Nanoscale surface tracking of laser material processing using phase shifting diffraction interferometry.* Optics Express, 2014. **22**(12): p. 14493-14504.

28. Ray, N.J., et al., *Birefringent Glass-Engraved Tilted Pillar Metasurfaces for High Power Laser Applications.* Advanced Science, 2023. **10**(24): p. 2301111.
22

# All-Glass Low Reflectance Metasurface Lenses Based on Laser Self-generated Nano-Particles – supplementary

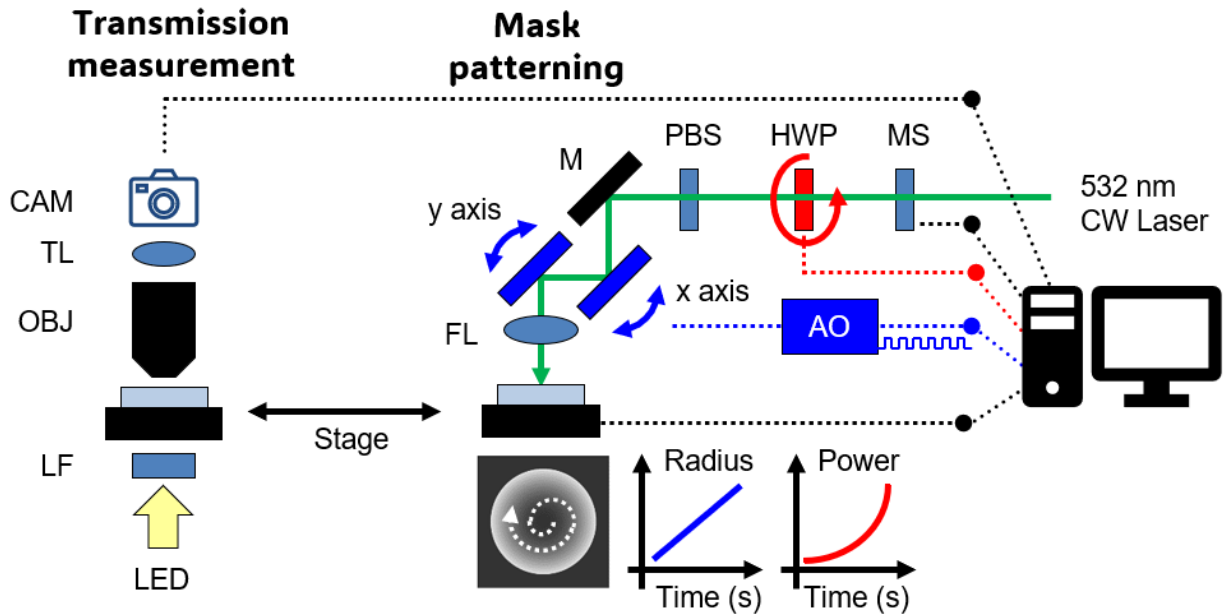

**Fig. S.1. Setup for obtaining prescribed transmission profile with radial laser scan:** setup description. MS: Mechanical shutter, HWP: Halfwave plate, PBS: Polarizing beam splitter, M: mirror, FL: focusing lens, CAM: camera, TL: tube lens, LF: line filter, OBJ: objective

Further description of the setup:

- CW 532 nm wavelength laser: 5W, Sprout-G, Lighthouse Photonics
- Two axis galvanometers: Thorlabs, GVS02
- Focusing lens of laser on sample: F-Theta lens, 15-185, Edmund, f=100 mm
- The 405 nm LED was equipped with a filter: 405 nm line filter with 10 nm FWHM
- Camera for the transmission measurement: PRIME95B, Photometrics
- Objective lens for camera: x2 Mitutoyo, NA: 0.055

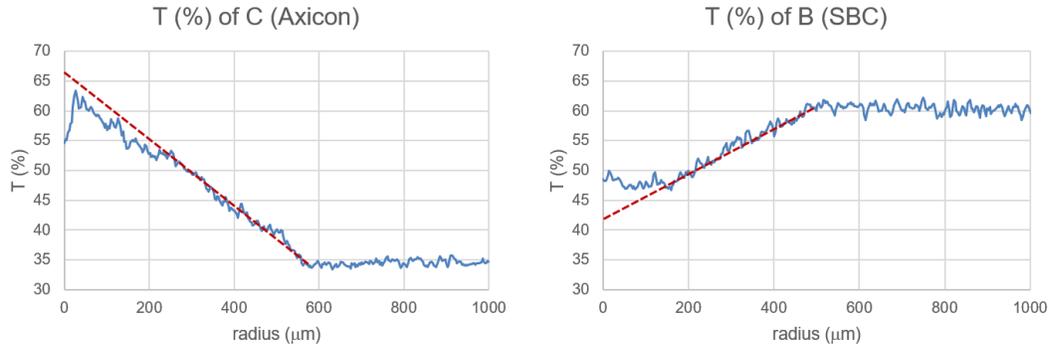

**Fig. S.2. Radial transmission profile of the nanoparticle masks used for the metaoptics: axicon and SCB**

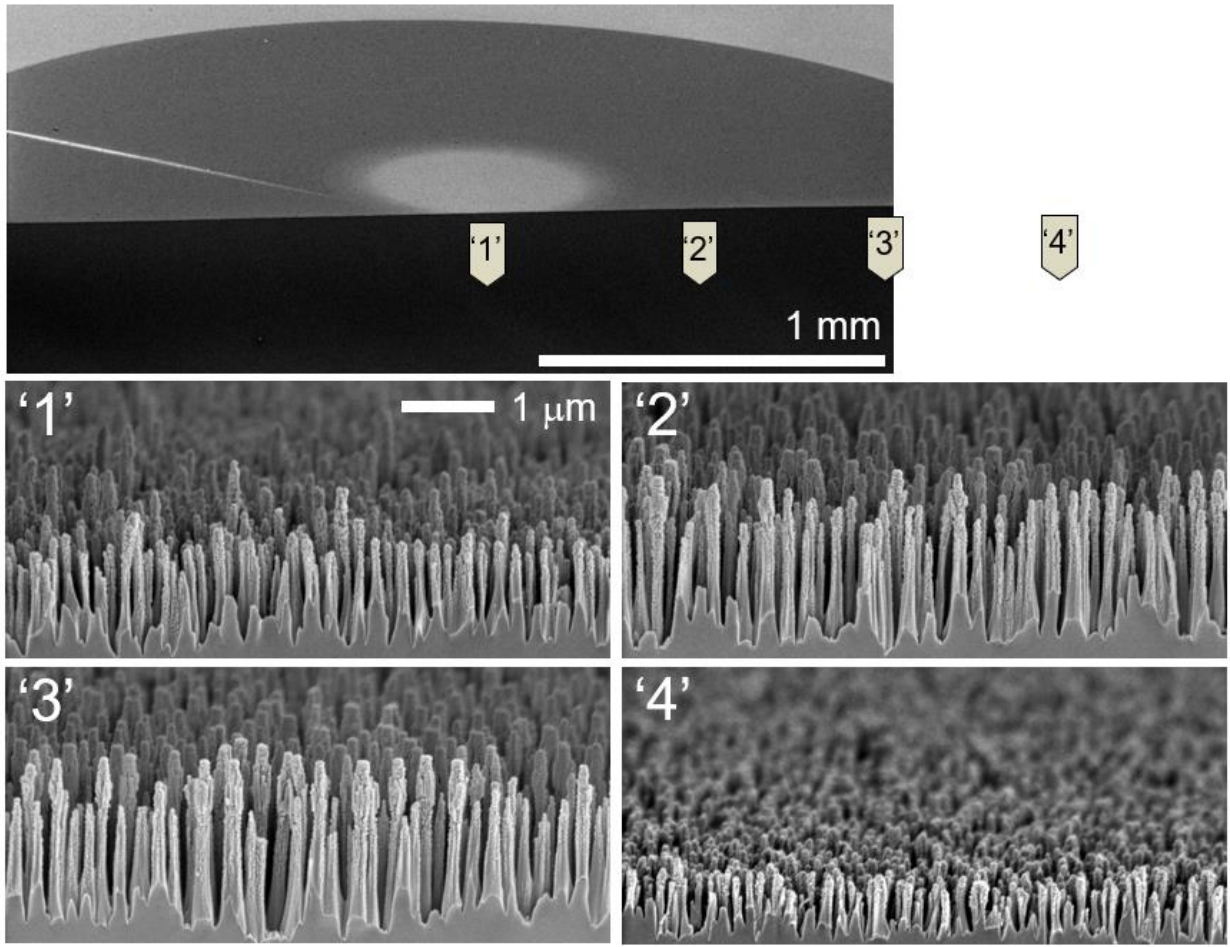

**Fig. S.3. Cross-sectional SEM of SCB metasurface:** cross-sectional SEM of the SCB MS sample after cleaving, at 4 segments – as indicated. The linear line on the MS is a scratch forming on it in the process of cleaving (not originally on the fabricated and performance measured metaoptics, as clear from the results in the paper)

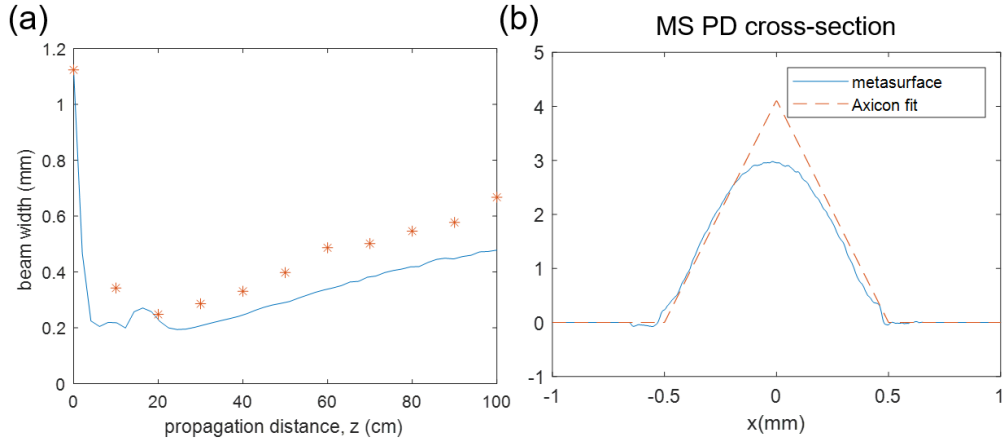

**Fig. S.4. Metasurface axicon beam propogation:** (a) Comparison of beam width (evaluated at half maximum) as a function of the beam propagation distance after the MS: simulation (solid blue) and measurements (red stars). The fabricated axicon is close to the idealized one performance yet has slightly shorter focal depth. (b) the PD cross-section of the MS and the idealized axicon fit.

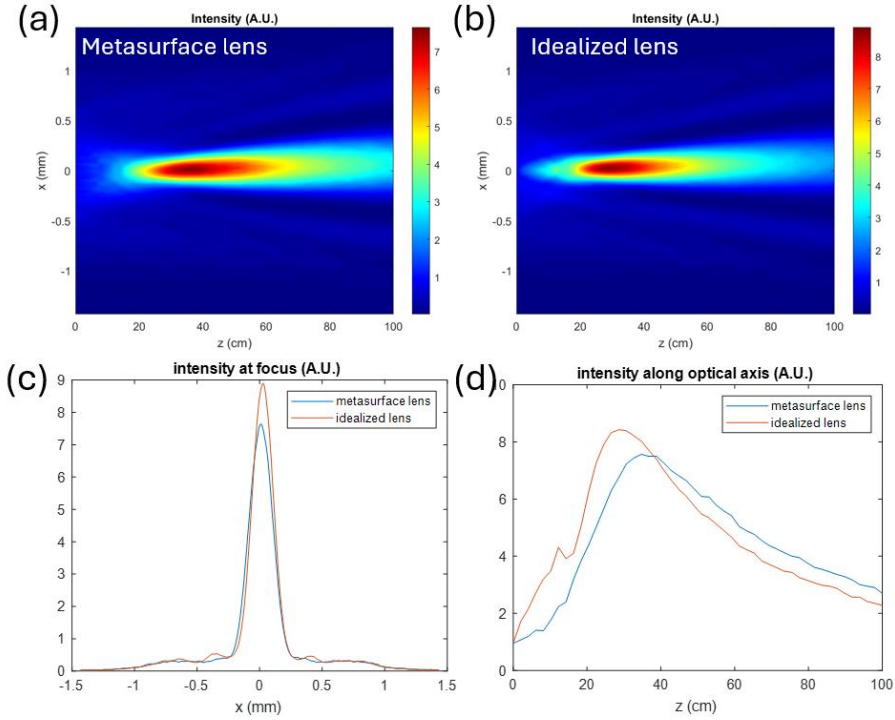

**Fig. S.5. Metasurface axicon beam propogation:** for the measured and idealized axion lens that, presented in Fig. S.4: the transversal intensity distribution for (a) measured and (b) idealized show very similar behavior. (c) The transvers intensity along x-axis at the peak intensity distance plane is presented comparing both lenses. The power in the main lobe for both is about 79.5% (The portion is calculated along the x-segment of between the idealized lens first nodes, as the ratio of sum on that segment divided by the total sum. The exact portion is slightly higher for the measured metasurface, 79.7% vs 79.3%, likely due to the center similarity to parabolic lens resulting in less power diverted to secondary bands slightly exceeding the power loss from the main lobe). (d) The axial intensity along the z-axis is compared, showing a similar elongated focal depth of about 6 cm, but slightly z-shifted.

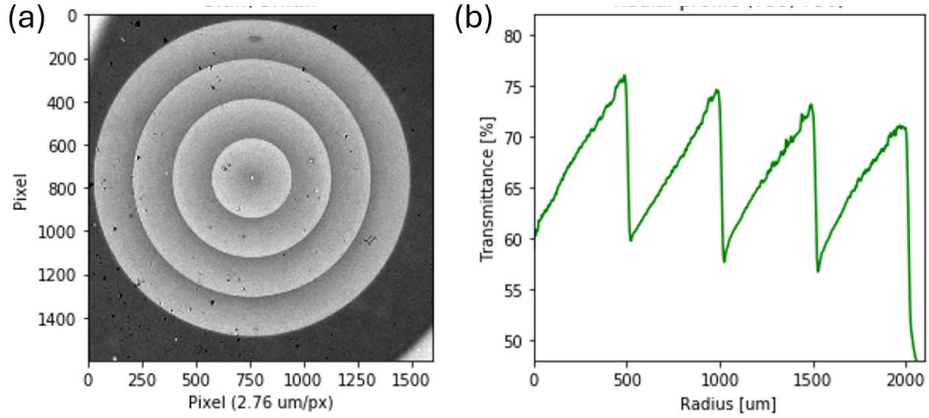

**Fig. S.6. Laser patterning of a diffractive optics-type shaped NP mask (unoptimized):** Transmission of a patterned mask with four Fresnel zones: (a) distribution and (b) radial line-out. The laser mask-patterning method is applied here using a radial power profile, P(r), that consists of four cycles of linearly increasing power from minimal to maximal power followed by a rapid decay to minimal power. This sawtooth P(r) shape is applied without the converging approach presented in the study or further optimization. Because the radial scan speed is constant, the dwell time decreases with the radius, resulting in an apparent reduction in the maximal transmission achieved in each of the four cycles as the radius increases. Nevertheless, the converging approach, if applied, is expected to adjust the power with the radius and compensate for this effect. This unoptimized demonstration is intended to demonstrate the ability to manufacture the sharp transitions needed for diffractive lens-type elements, a potential topic for future study.